\documentclass[iicol]{sn-jnl}% Default with double column layout

\usepackage{graphicx}%
\usepackage{multirow}%
\usepackage{amsmath,amssymb,amsfonts}%
\usepackage{amsthm}%
\usepackage{mathrsfs}%
\usepackage[title]{appendix}%
\usepackage{xcolor}%
\usepackage{textcomp}%
\usepackage{manyfoot}%
\usepackage{booktabs}%
\usepackage{algorithm}%
\usepackage{algorithmicx}%
\usepackage{algpseudocode}%
\usepackage{listings}%

\usepackage{diffcoeff}
\usepackage{mathtools}
\usepackage{xspace}
\usepackage{tabularx}
\usepackage{bbold}

\DeclarePairedDelimiterX{\braket}[2]{\langle}{\rangle}{#1\,\delimsize\vert\,\mathopen{}#2}
\DeclarePairedDelimiterX{\ketbra}[2]{\lvert}{\rvert}{#1\,\delimsize\rangle\mathopen{}\delimsize\langle\,\mathopen{}#2}
\DeclarePairedDelimiterX{\proj}[1]{\lvert}{\rvert}{#1\,\delimsize\rangle\mathopen{}\delimsize\langle\,\mathopen{}#1}
\DeclarePairedDelimiterX{\mel}[3]{\langle}{\rangle}{#1\,\delimsize\vert\mathopen{}#2\delimsize\vert\,\mathopen{}#3}

\DeclarePairedDelimiterX{\Braket}[2]{{\bm{(}}}{{\bm{)}}}{#1\delimsize\bm{\vert}\mathopen{}#2}
\DeclarePairedDelimiterX{\Ketbra}[2]{{\bm{|}}}{{\bm{|}}}{#1\delimsize\bm{)}\mathopen{}\delimsize\bm{(}\mathopen{}#2}

\DeclarePairedDelimiterXPP{\Srel}[2]{S}{(}{)}{}{#1\delimsize\|\mathopen{}#2}
\newcommand{\Tr}{\text{Tr}}

\newcommand{\R}{\text{R}}
\renewcommand{\L}{\text{L}}

\newcommand{\Ome}{\Omega_\epsilon}
\newcommand{\OmL}{\Omega_\Lambda}
\newcommand*{\Tp}{\tau}

\newcommand{\Ct}{\text{C}}
\newcommand{\dT}{\delta T}

\newcommand*{\etaC}{\eta^\Ct}

\newcommand{\Work}{\mathcal{W}}
\newcommand{\QR}{\mathcal{Q}_{\R}}
\newcommand{\QL}{\mathcal{Q}_{\L}}
\newcommand{\beps}{\bar{\epsilon}}
\newcommand{\bepsopt}{\bar{\epsilon}_\text{opt}}
\newcommand{\deps}{\delta\epsilon}

\begin{document}
    \title[]{Non-geometric pumping effects on the performance of interacting quantum-dot heat engines}

    \author*[]{\fnm{Juliette} \sur{Monsel$^{1*}$}}
    \email{monsel@chalmers.se}

    \author[]{\fnm{Jens} \sur{Schulenborg$^{1,2}$}}

    \author[]{\fnm{Janine} \sur{Splettstoesser$^1$}}

    \affil[1]{\orgdiv{Department of Microtechnology and Nanoscience (MC2)}, \orgname{Chalmers University of Technology}, \orgaddress{\postcode{S-412 96} \city{G\"oteborg}, \country{Sweden}}}
    \affil[2]{\orgdiv{Center for Quantum Devices}, \orgname{Niels Bohr Institute, University of Copenhagen}, \orgaddress{\postcode{DK-2100} \city{Copenhagen}, \country{Denmark}}}

    \abstract{Periodically driven quantum dots can act as counterparts of cyclic thermal machines at the nanoscale. In the slow-driving regime of geometric pumping, such machines have been shown to operate in analogy to a Carnot cycle. For larger driving frequencies, which are required in order to increase the cooling power, the efficiency of the operation decreases. Up to which frequency a close-to-optimal performance is still possible depends on the magnitude and sign of on-site electron-electron interaction. Extending our previous detailed study on cyclic quantum-dot refrigerators [Phys. Rev. B \textbf{106}, 035405 (2022)], we here find that the optimal cooling power remains constant up to weak interaction strength compared to the cold-bath temperature. By contrast, the work cost depends on the interaction via the dot’s charge relaxation rate, as the latter sets the typical driving frequency for the onset of non-geometric pumping contributions.}

    \keywords{quantum transport, quantum thermodynamics, quantum dots}

    \maketitle

    %%=============================================================%%
    \section{Introduction}
    %%=============================================================%%

\begin{figure}[tb]
        \includegraphics[width=\linewidth]{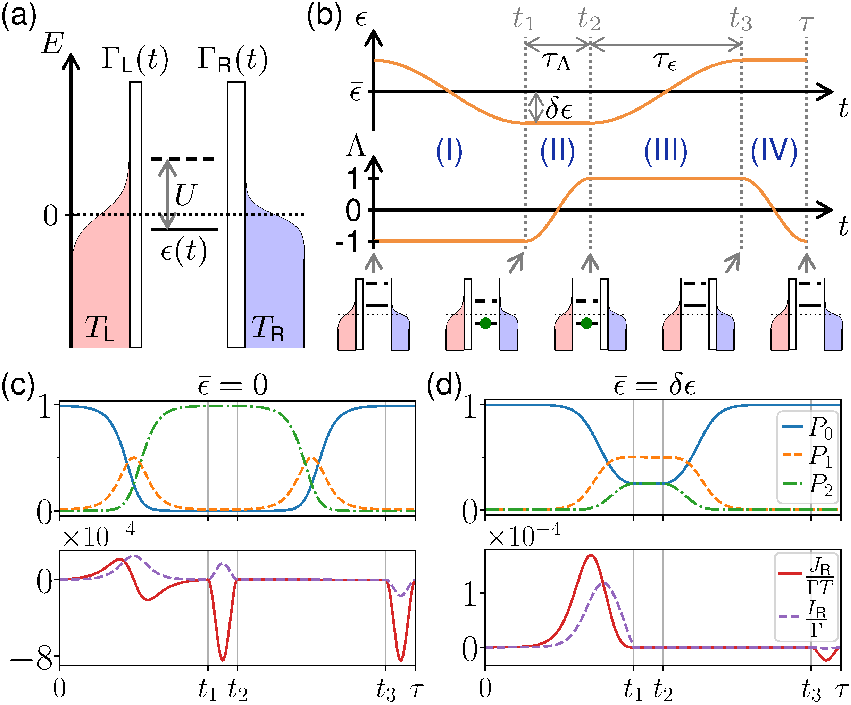}
        \vskip-1mm
        \caption{\label{fig1}
            (a) Sketch of the quantum dot model, see text for details about the parameters.
            (b) 4-stroke driving protocol: $\epsilon$ and $\Lambda$ as functions of time over one driving cycle, see Eq.~\eqref{driving}. The schematics indicate the state of the dot at the start of each stroke for $\beps = 0$ and $\deps \gg T$.
            (c,d)~Evolution of the populations, $P_i$,  and currents, $I_\mathrm{R},J_\mathrm{R}$ during one cycle for (c) $\beps = 0$ and (d) $\beps = \deps$.
            Furthermore, $U = 0$, $\deps/T = 5$, $\Gamma/T = 10^{-2}$, $\dT/T = 10^{-2}$, $\Ome/\Gamma = 10^{-4}$ and $\OmL = 5\Ome$.
        }
    \end{figure}

    Cyclic thermal machines have been widely studied in nanoscale systems \cite{Alicki1979May,Feldmann2003Jul,Allahverdyan2008Apr,Kosloff2014Apr,Kosloff2017Mar,Myers2022Jan}. One possibility to design such a thermal machine at the nanoscale is via periodic driving---pumping---of quantum dots~\cite{Juergens2013Jun,Hino2021Feb,Monsel2022Jul}, see Fig.~\ref{fig1}(a) and (b).
    In these quantum-dot realizations of heat engines, electron-electron interaction can play an important role in the performance. This is true for the standard repulsive Coulomb interaction, but strong \textit{attractive} interaction has also been analyzed in pumping through quantum dots~\cite{Placke2018Aug} and for steady-state thermoelectric systems~\cite{Schulenborg2020Jun}.
    The most basic setup allowing for such a nano-electronic implementation of a cyclic thermal machine is a driven single-level quantum dot coupled to two electronic baths.
    For this basic setup, we have recently identified the key physical effects and, in particular, the role of many-body on-site interaction, on the thermodynamic performance~\cite{Monsel2022Jul}. We showed that by periodically driving the couplings to the hot and cold baths and the energy of level of the dot, the system can be operated as a refrigerator or a heat pump and the sign of the interaction, namely strong attractive or strong repulsive electron-electron interaction, has an important impact on the machine performance. However, this previous work focused on the geometric approach arising from slow sinusoidal driving and small driving amplitudes which are not the most typical conditions for cyclic heat engines.

    The present work extends our previous study of a cyclic quantum-dot refrigerator to a more traditional four-stroke thermodynamic cycle, as depicted in Fig.~\ref{fig1}(b), and to driving frequencies beyond the adiabatic-response regime of the pump. The latter frequencies are of interest to increase the cooling power of the refrigerator, but at the same time, the efficiency is known to be the largest when the frequency is small (Carnot limit). We study in detail the role of the many-body interaction (weak to strong repulsive or attractive) on the upper limit of the driving frequency that still allows for a close to Carnot performance. We consider  weak to strong repulsive and attractive electron-electron interaction.

    %%=============================================================%%
    %%=============================================================%%

    %%=============================================================%%
    \section{Heat transport induced by a driven quantum dot}
    %%=============================================================%%

    We consider a thermal machine consisting of a single-level spin-degenerate quantum dot as working substance, tunnel-coupled to two electronic baths (reservoirs), as depicted in Fig.~\ref{fig1}(a). The quantum-dot Hamiltonian reads
       ${H}=\epsilon{N}+U{N}_\uparrow{N}_\downarrow$,
    with $\epsilon$ the energy of the dot level, $U$ the interaction energy, ${N}_\sigma$ the number of electrons of spin $\sigma=\uparrow,\downarrow$ in the dot and $N = {N}_\uparrow + {N}_\downarrow$. Note that the two-particle interaction $U$ is \emph{not} treated perturbatively, and can take on arbitrary positive (repulsive Coulomb interaction) or negative (attractive interaction) values.
    The energy-independent coupling strengths between the dot and reservoirs are $\Gamma_{\L/\R} = \Gamma (1 \pm \Lambda)/2$, where $\Gamma$ is a fixed tunnel coupling strength and $\Lambda$ the left-right asymmetry.
    Each reservoir $\alpha=\L,\R$ is assumed to be an effectively noninteracting, spin-degenerate metallic lead in equilibrium at temperature $T_\alpha$ and electrochemical potential $\mu_\alpha$. The average particle (+)/hole (-) occupation number at energy $E$ is given by the Fermi distribution $f^\pm_\alpha(E) = \left(1 + \exp(\pm(E-\mu_\alpha)/k_\text{B} T_\alpha)\right)^{-1}$.  In the following, we adopt the convention $\hbar \equiv k_\text{B} \equiv e \equiv 1$ and use the temperature $T$ as our reference energy scale.

    We are interested in operating this quantum dot as a cyclic refrigerator. We therefore choose identical chemical potentials for both reservoirs and use them as the zero of the energy scale, namely $\mu_\R = \mu_\L \equiv 0$. We consider the left reservoir to be hotter than the right one and write $T_\R \equiv T$ and $T_\L \equiv T + \dT$, with $\dT > 0$.
    The thermal machine is operated with the following four-stroke protocol: we alternatively modulate the level energy $\epsilon$ and the tunneling asymmetry $\Lambda$ with piecewise cosine functions, see Fig.~\ref{fig1}(b), at respective frequencies $\Ome = \pi/\tau_\epsilon$ and $\OmL = \pi/\tau_\Lambda$,
          \begin{subequations}\label{driving}
                \begin{align}
            \epsilon(t) &= \left\{
            \begin{array}{lc}
                \bar{\epsilon} + \delta\epsilon \cos(\Ome t) & \text{(I)}\\
                \bar{\epsilon} - \delta\epsilon  & \text{(II)}\\
                \bar{\epsilon} - \delta\epsilon\cos(\Ome (t - t_2))  & \text{(III)} \\
                \bar{\epsilon} + \delta\epsilon  & \text{(IV)}\\
            \end{array} \right., \\
            \Lambda(t) &= \left\{
            \begin{array}{lc}
                -1 & \text{ (I)}\\
                -\cos(\OmL (t - t_1)) \hspace*{0.4cm} & \text{ (II)}\\
                1  & \text{ (III)}\\
                \cos(\OmL (t - t_3)) & \text{ (IV)}
            \end{array} \right.,
        \end{align}
       \end{subequations}
    with $t_1 = \tau_\epsilon$, $t_2 = t_1 + \tau_\Lambda$, $t_3 = t_2 + \tau_\epsilon$ and the total cycle duration $\Tp = 2\tau_\epsilon + 2\tau_\Lambda$. The level energy is modulated with amplitude $\deps$ around a chosen average value $\beps$ while the tunneling asymmetry is always modulated around 0 with amplitude 1 so that the dot is only in contact with the cold bath during stroke (I) and only with the hot bath during stroke (III). The bath couplings are inverted during strokes (II) and (IV).  This procedure differs from our previous analysis \cite{Monsel2022Jul}, where we chose a simple and continuous cosine-shaped driving both for the level position and the coupling asymmetry.

    We describe the dynamics of such a thermal machine by a master equation~\cite{Splettstoesser2006Aug,Cavaliere2009Sep,Reckermann2010Jun} $\partial_t \rho = W \rho$ for the reduced density operator of the dot $\rho$; this is valid in the weak-coupling regime, $\Gamma \ll T$, and for moderately slow driving, $0 < \Ome \deps/T,\OmL \lesssim \Gamma$. The kernel ${W}$ acting on $\rho$ is here time dependent due to the modulation of $\epsilon$ and $\Lambda$ during the strokes.
    Furthermore, only the diagonal elements of $\rho$, namely the probabilities $P_i$ of the occupation states $i = 0,1,2$ of the dot, are of relevance. The kernel $W$ gives the transition rates between those occupation states and can be split into the separate contributions of each reservoir, ${W} = {W}^\L + {W}^\R$.  The charge and heat currents from reservoir $\alpha$ into the dot read $I_{\alpha}(t) = \Tr[N (W^\alpha \rho(t))]$ and $J_{\alpha}(t) = \Tr[H (W^\alpha \rho(t))]$~\cite{Schulenborg2017Dec}. The transition rates can be computed from Fermi's golden rule and are proportional to $\Gamma_\alpha$ times a Fermi function. See Ref.~\cite{Monsel2022Jul} for details about the model and its solution, exploiting a dissipative symmetry of the master equation kernel, coined \emph{fermionic duality}.
     %%=============================================================%%
    \begin{table*}[tb]
        \begin{center}
            \begin{minipage}{\textwidth}
                \caption{\label{table}
                    Reference numerical values for different interaction strengths, from strongly attractive to strongly repulsive. The parameters are the same as in Fig.~\ref{fig2}.
                }
                \begin{tabular*}{\textwidth}{@{\extracolsep{\fill}}lccccc@{\extracolsep{\fill}}}
                    \toprule
                    $U/T$ & -10 & -0.5 & 0 & 0.5 & 10 \\
                    \midrule
                    $\bepsopt/T$ & $10.0$ & $5.25$ & $5.00$ & $4.75$ & $5.00$\\
                    $\QR^\Ct/T$ &$0.733$ & $1.378$ & $1.385$ & $1.377$ & $1.098$\\
                    max. $\Ome/\Gamma$ s.t. $\QR/\QR^\Ct > 90 \%$&  $6.47 \times 10^{-4}$ & $4.66 \times 10^{-2}$ & $4.85 \times 10^{-2}$ & $4.90 \times 10^{-2}$ & $6.15 \times 10^{-2}$\\
                    max. $\Ome/\Gamma$ s.t. $\eta/\etaC > 90 \%$ & $3.41 \times 10^{-6}$ & $2.40 \times 10^{-4}$ & $2.52 \times 10^{-4}$ & $2.56 \times 10^{-4}$ & $3.18 \times 10^{-4}$\\
                    \botrule
                \end{tabular*}
            \end{minipage}
        \end{center}
    \end{table*}
    %%=============================================================%%

    In order to quantify the performance of the driven dot as refrigerator, like in Ref.~\cite{Monsel2022Jul}, we are specifically interested in the heat extracted from the cold bath, i.e. the right reservoir, per cycle as well as the work $\Work$ supplied by the driving,
    \begin{equation}
        \QR = \int_0^{\Tp} \dl tJ_\R(t),\;\;\Work = \int_0^{\Tp} \dl t \Tr\left[\diff{{H}}{t}\rho \right].
    \end{equation}
    Indeed, since $\mu_\R = 0$, the heat current is equal to the energy current and, in the weak-coupling regime, the modulation of the tunnel barriers does not cost any work. Finally, we define the coefficient of performance $\eta = \QR/\Work$. In the corresponding ideal Carnot cycle, that is for $\Ome\to 0$ and $\OmL \to \infty$, there is no heat exchanged during strokes (II) and (IV), while the second law of thermodynamics for strokes (I) and (III) gives $\QR^\Ct = T \Delta S^\text{(I)}$ and $\QL^\Ct = (T + \dT)\Delta S^\text{(III)}$. $\Delta S^\text{(X)}$ denotes the change in the von Neumann entropy of the dot, $S(\rho) = -\Tr[\rho \log\rho]$, during stroke (X). Finally, applying the first law to the full cycle and using $ \Delta S^\text{(III)} = -\Delta S^\text{(I)}$, we obtain the work cost of the Carnot cycle $\Work^\Ct = \dT \Delta S^\text{(I)}$ and the expected Carnot efficiency $\etaC = T/\dT$.

    %%=============================================================%%
    \section{Optimal operation constrained by interaction}
    %%=============================================================%%

    The goal of the refrigerator is to perform heat extraction from the cold bath during stroke (I), while avoiding to have unwanted heat flows during strokes (II) and (IV).
    To highlight the impact of interaction on the optimal performance, we first introduce the operation principle for a noninteracting quantum dot, $U=0$,  modulated with an amplitude $\deps$ that is large compared to temperature.
    The system dynamics during a driving cycle---here always chosen as depicted in Fig.~\ref{fig1}(b)---are shown in Fig.~\ref{fig1}(c) for driving around  $\beps = 0$ and for $\beps = \deps$ in panel (d). In the former case, the dot occupation increases from 0 to 2 when lowering $\epsilon$ during stroke (I) since the transition energy $\epsilon(0) = \deps \gg T$ is way above the cold reservoir electrochemical potential, and $\epsilon(t_1) = -\deps \ll -T$ way below.
    This cycle is not favorable since the desired heat flow during stroke (I) averages to zero, giving a total $\QR < 0$. The evolution is different for $\beps = \deps$ since at the end of stroke (I), the dot level is exactly at resonance with the reservoir, leading to $\QR > 0$.
    This shows that it is favorable if the addition energies approach, but do not cross the common electrochemical potential of the two reservoirs, such that transport during the whole stroke is due to electrons only (respectively due to holes only for a cycle starting below the Fermi energy). Furthermore, there is no leakage heat current when the coupling to the baths is modified, if the addition energy is exactly at resonance during one of the strokes (II) or (IV).

    Importantly, in the case $\beps=\deps$, a small repulsive interaction, $U > 0$, would give a similar evolution but with a smaller probability of double occupation, leading to a reduced dot filling during the cycle. Conversely, a small attractive interaction, $U < 0$, gives a larger probability of double occupation, concomitant with a slight crossing of one of the addition energies with the zero electrochemical potentials. In other words, the optimal refrigerator level working point $\beps$ in terms of cooling power is generally shifted away from $\beps=\deps$ for sizable local interaction strength. Then, for $U \gg T$, only one of the two addition energies is involved in the cycle, so the reasoning for $U = 0$ still applies to either the $0\to1$ or $1\to 2$ transition. Conversely, the case $-U \gg T$ is special because the optimal working point relies on the \emph{pair resonance} $\epsilon\approx -U/2$ ---induced by strong attractive interaction--- to coincide with the electrochemical potential at the end of stroke (I).

    The following analysis will always assume a $\beps = \bepsopt$ maximizing $\QR^\Ct = T \Delta S^\text{(I)}$ by, ideally, inducing a transition from a pure to a maximally mixed dot state during stroke (I). 
    For $U \lesssim T$, $\bepsopt = \deps - U/2$ but, for $U \gg T$, the two addition energies can be separated and $\bepsopt \simeq \deps$, see the first row in Table~\ref{table}. The Carnot limit $\QR^\Ct$ of the extracted heat at these optimal points is the largest for the non-interacting quantum dot as described above; attractive interaction reduces the value more strongly than repulsive because the $0\rightarrow 1$ transition can only be thermally activated at the pair resonance.

    As clear from Figs.~\ref{fig1}(c,d), it is not only important to maximize the heat extraction during stroke (I), but also to avoid heat leakage during strokes (II) and (IV). Apart from an optimal working point, this leakage is reduced by shorter stroke durations. Since strokes (II) and (IV) are not dissipative in the weak-coupling regime, we thus set a relatively large frequency $\OmL = 2\Gamma$.

\begin{figure}[hb]
        \includegraphics[width=\linewidth]{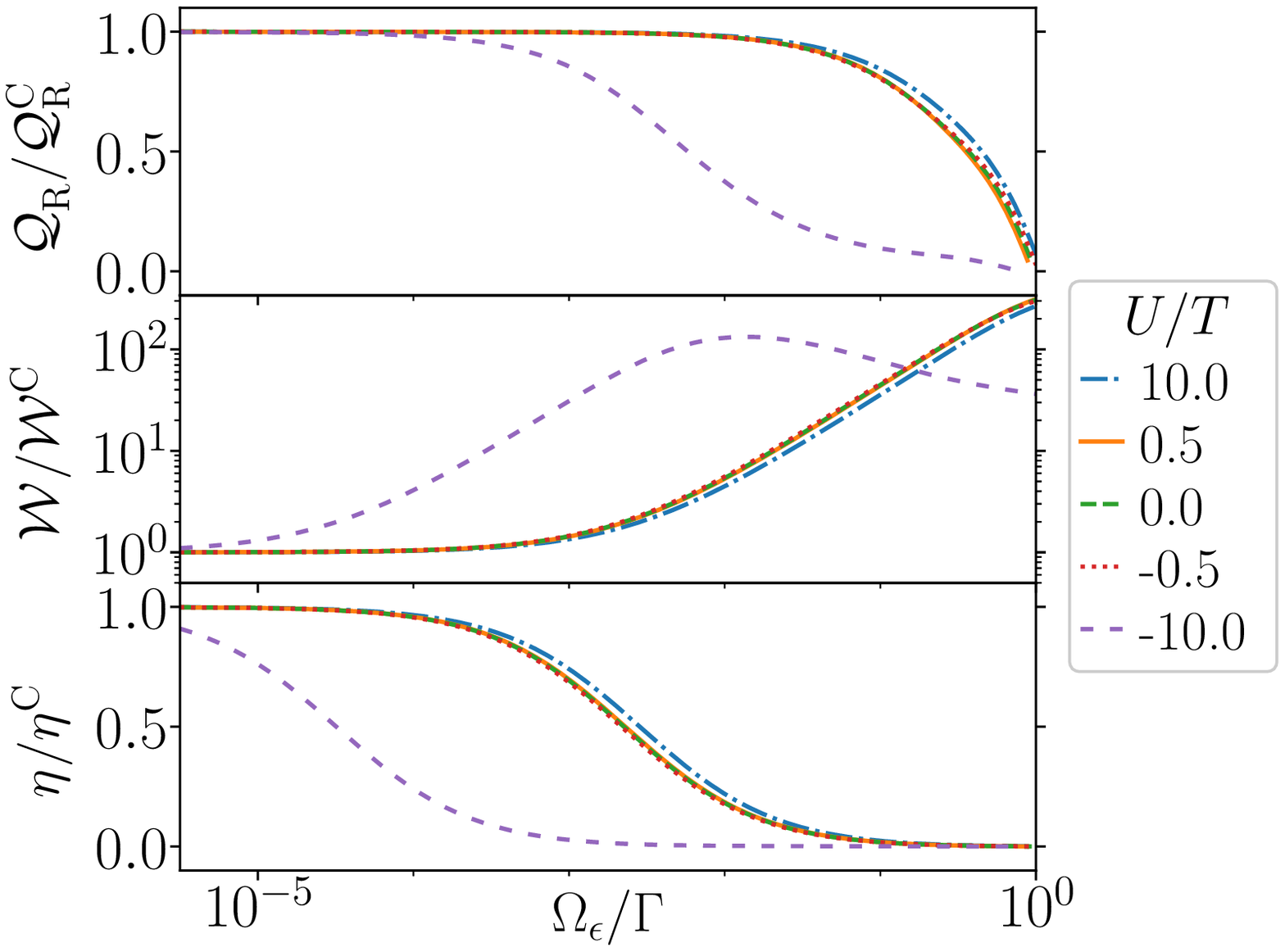}
        \vskip-2mm
        \caption{\label{fig2}
            Extracted heat per cycle $\QR$, work cost $\Work$ and efficiency $\eta$ divided by the corresponding quantities in the Carnot limit as functions of $\Ome$ for different  interaction strengths $U$. The other parameters are $\Gamma/T = 10^{-2}$, $\dT/T = 10^{-2}$, $\deps/T = 5$, $\beps = \bepsopt$ and $\OmL = 2\Gamma$.
        }
    \end{figure}

    While the Carnot values for heat extraction and efficiencies can be obtained for infinitely small driving frequencies, reasonably large frequencies are required to reach measurable cooling power (heat extracted per unit of time). It is therefore important to identify up to which driving frequency $\Ome$ the refrigerator supports either large heat extraction, e.g., $\QR/\QR^\Ct > 90 \%$, and/or large efficiencies, e.g., $\eta/\etaC > 90 \%$. These operation frequency bounds are directly tied to the magnitude and sign of on-site electron-electron interaction, as shown in the third and fourth row of Table~\ref{table}, and in Fig.~\ref{fig2}, and as we will discuss in the following.

    %%=============================================================%%
    %%=============================================================%%

    Fig.~\ref{fig2} shows the extracted heat per cycle $\QR$, the work cost $\Work$ and the efficiency $\eta$ as function of $\Ome$ for interaction strengths ranging from strongly attractive to strongly repulsive. All cases converge towards the ideal Carnot cycle values for infinitely slow $\Ome \to 0$, and all the curves are very similar for strong repulsion, $U = 10T$, up to weak attraction, $U = -0.5T$. The convergence towards the Carnot limit is slightly faster for strongly repulsive interaction. On the contrary, the $\QR(\Ome)$ vary substantially in shape as a function of $U$ approaching strong attraction ($U = -10T$) , with significantly slower convergence towards the Carnot limit.
    This is because the relevant rate to compare to $\Ome$ is the interaction-dependent charge rate~\cite{Splettstoesser2010Apr,Monsel2022Jul}
    \begin{equation}
        \gamma_\alpha^c=\Gamma[f_\alpha^+(\epsilon(t))+f_\alpha^-(\epsilon(t) + U)],
    \end{equation}
    with $\alpha = \R$ during stroke (I) and $\alpha = \L$ during stroke (III). While this charge rate is bounded by $\Gamma_\alpha\leq\gamma_\alpha^c\leq2\Gamma$ for $U \geq 0$~\cite{Schulenborg2016Feb}, it approaches $2\Gamma$ in the single-occupation regime enabled by repulsive interaction. This allows for faster driving frequencies in protocols ending up in precisely this singly occupied state. However, attractive interaction $U < 0$ can suppress $\gamma_\alpha^c$ below $\Gamma_\alpha$. Namely, while $\gamma_\alpha^c \approx\Gamma_\alpha$ still holds during the whole stroke for weak $U\!\sim\! -T$, this is not the case in the strongly attractive regime: around $\bepsopt - \deps$, that is, for a tendentially singly occupied dot at the end of stroke (I) and the beginning of stroke (III), $\gamma_\alpha^c$ becomes vanishingly small for $- U \gg T$. This, in turn, strongly bounds the frequency at which the refrigerator can operate close to the Carnot limit.
    The modulation-frequency dependence of the work spent for the refrigerator operation in the strongly attractive regime also reflects this, as already small frequencies $\Omega\leq 10^{-1}\Gamma$  no longer allow the dot to be fully charged and discharged during the cycle. In contrast, repulsive interaction may thereby in fact benefit $\eta$, as the larger charge relaxation rate can result in a lower $\Work$.

    %%=============================================================%%
    \begin{figure}[tb]
        \includegraphics[width=\linewidth]{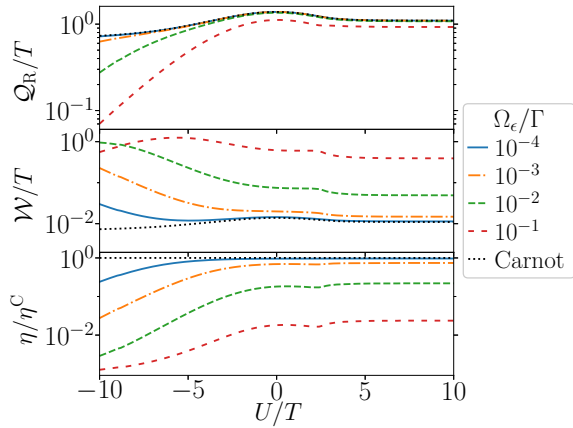}
        \vskip-2mm
        \caption{\label{fig3}
        Extracted heat per cycle $\QR$, work cost $\Work$ and efficiency $\eta$ as functions of the interaction strength $U$ for different driving frequency $\Ome$. The other parameters are the same as in Fig.~\ref{fig2}. The black dotted lines correspond to the case of the ideal Carnot cycle.
        }
    \end{figure}
    %%=============================================================%%

    Complementary to Fig.~\ref{fig2}, we plot the different thermodynamic quantities as functions of the interaction strength $U$ in Fig.~\ref{fig3} for different driving frequencies $\Ome$.
    Again, while the performance does not strongly depend on interaction as long as $U > 0$, the deviation from the ideal Carnot cycle gets stronger the more negative $U$ becomes.

    Additionally, we observe a feature in all three plotted quantities of Fig.~\ref{fig3} around $U/T \simeq 2.6$. It appears at the point where $U$ becomes sufficiently large compared to temperature to separate the two addition energies $\epsilon$ and $\epsilon + U$; note that the broadening of the Fermi function is given by $4k_\mathrm{B}T$.  This separation of addition energies influences the amount of accessible states (degeneracy) at different times during the strokes and thereby directly affects the dot entropy $S(\rho)$. This in turn impacts the extracted heat~\cite{Juergens2013Jun}: concretely, for our choice of $\deps$ and $\beps$, we have $\QR^\Ct \simeq S(\rho(t_1))$.
    The impact of the degeneracy on thermodynamic observables has previously been shown in quantum-dot experiments~\cite{Hofmann2016Nov,Hartman2018Nov,Kleeorin2019Dec}, and we here show its features in the operation of a cyclic thermal machine.

    %%=============================================================%%
    \section{Conclusion}
    %%=============================================================%%
    This manuscript has complemented the detailed analysis~\cite{Monsel2022Jul} of a driven quantum dot operated as a cyclic refrigerator with a study of how the dot-local interaction impacts the performance of such a refrigerator when operated close to a standard Carnot cycle. We have therefore chosen a convenient four-stroke modulation, alternating between a driven quantum dot level and system-bath coupling.
    Generally, the Carnot limit of performance requires adiabatic engine strokes.
    Here, we have quantified how the interaction bounds the highest frequency still allowing the system to operate \emph{close} to this limit. We have furthermore identified significant differences between repulsive and (effectively) attractive interaction. These differences primarily originate from the fact that while the charge relaxation rate of the dot is always of the order of the tunnel barrier transparency $\Gamma$ for repulsive interaction, the charge rate may almost vanish for effective on-site electron-electron attraction. Also the Carnot value of the heat extracted by the refrigerator is reduced in the case of strong attractive interaction, since thermal activation of tunneling processes is required in the optimal operation regime.
    Finally, we have discussed degeneracy effects ---visible only in the presence of electron-electron interaction--- that manifest as a distinct feature in the extracted heat, work cost and the efficiency of the cyclically operated quantum-dot refrigerator.

    %%=============================================================%%
    \backmatter
    %%=============================================================%%

    %%=============================================================%%
    \bmhead{Acknowledgments}
    %%=============================================================%%

    This work has received funding from the European Union's H2020 research and innovation program under grant agreement No.~862683. We also acknowledge funding from The Knut and Alice Wallenberg foundation and from the Swedish Vetenskapsr\r{a}det (project number 2018-05061).

\end{document}